\pdfoutput=1
\RequirePackage{ifpdf}
\ifpdf 
\documentclass[pdftex]{sigma}
\else
\documentclass{sigma}
\fi

\DeclareMathAlphabet{\mathpzc}{OT1}{pzc}{m}{it}

\def\e{{\rm e}}

\numberwithin{equation}{section}

\begin{document}

\allowdisplaybreaks

\renewcommand{\thefootnote}{$\star$}

\renewcommand{\PaperNumber}{031}

\FirstPageHeading

\ShortArticleName{Superintegrable St\"ackel Systems on the Plane: Elliptic and Parabolic Coordinates}

\ArticleName{Superintegrable St\"ackel Systems on the Plane:\\ Elliptic and Parabolic Coordinates\footnote{This
paper is a contribution to the Special Issue ``Superintegrability, Exact Solvability, and Special Functions''. The full collection is available at \href{http://www.emis.de/journals/SIGMA/SESSF2012.html}{http://www.emis.de/journals/SIGMA/SESSF2012.html}}}

\Author{Andrey V.~TSIGANOV}

\AuthorNameForHeading{A.V.~Tsiganov}

\Address{St.~Petersburg State University, St.~Petersburg, Russia}
\Email{\href{mailto:ndrey.tsiganov@gmail.com}{andrey.tsiganov@gmail.com}}

\ArticleDates{Received April 10, 2012, in f\/inal form May 21, 2012; Published online May 25, 2012}

\Abstract{Recently we proposed a generic construction of the additional integrals of motion for  the St\"ackel  systems applying  addition theorems to  the angle variables.  In this note we show some trivial examples associated with angle variables for elliptic and parabolic coordinate systems on the plane.}


\Classification{37J35; 70H06} 

\section{Introduction}
In classical mechanics  Hamiltonian system on a $2n$-dimensional phase space $M$ is called completely integrable in Liouville's sense if it possesses $n$ functionally independent integrals
of motion $H_1,\dots,H_n$ in  involution:
\[\dfrac{dH_i}{dt}=\{H,H_i\}=0,\qquad \{H_i, H_j\}= 0,\qquad i,j=1,\ldots,n,\]
where $H=H_1$ is the Hamilton function and $\{\cdot,\cdot\}$ is the Poisson bracket on $M$.

Superintegrable system is a system that is integrable in the Liouville sense and that possesses more functionally independent integrals of motion than degrees of freedom.  The construction of superintegrable St\"ackel systems using angle variables~$\omega_k$ has been proposed in \cite{ts08a, ts09e,ts08m,ts10r}.

In generic case the action variables $\omega_k$ are multi-valued  functions on the whole phase space~$M$.  In fact,  we can extract polynomial integrals of motion  from  angle variables only when we can apply addition theorems to the corresponding  Abelian integrals. As there are only few addition theorems for the Abel equations \cite{bak97, eul68} we can easily classify the corresponding superintegrable systems, see \cite{ts08a, ts09e,ts08m, ts10r}.

The goal of this brief note is to present  some trivial examples of applying this generic theory associated with  elliptic and parabolic coordinate systems on the plane. Superintegrable systems separable  in spherical coordinates can be found in~\cite{bmk09,km12}. The corresponding addition integrals of motion are  related with an addition theorem for the logarithmic angle variables. Of course, there is a trivial  generalization of the proposed method for all the orthogonal coordinate systems in~$\mathbb R^3$ (ellipsoidal, paraboloidal, cylindrical, prolate and oblate spheroidal coordinates etc).

The non-St\"ackel superintegrable systems in classical and quantum mechanics have been considered in~\cite{ts11,w11}. In contrast with the St\"ackel case we do not  have  a generic theory for  constructing  such superintegrable systems.

\section{The St\"{a}ckel systems}

The system associated with the name of St\"{a}ckel \cite{st95} is a holonomic system on the phase space~$\mathbb R^{2n}$, with the canonical variables
$q=(q_1,\ldots,q_n)$ and $p=(p_1,\ldots,p_n)$:
\begin{gather*} \Omega=\sum_{j=1}^n dp_j\wedge dq_j ,\qquad
\{p_j,q_k\}=\delta_{jk} .
\end{gather*}
The nondegenerate $n\times n$ St\"{a}ckel matrix $S$, whose $j$ column depends only on coordinate $q_j$, def\/ines $n$ functionally independent integrals of motion
\begin{gather*}
H_k=\sum_{j=1}^n \big( S^{-1}\big)_{jk}\bigl(p_j^2+V_j(q_j)\bigr) ,
\end{gather*}
 the separated relations
\begin{gather*}
p_j^2=
\sum_{k=1}^n H_k S_{kj}(q_j)-V_j(q_j) ,
\end{gather*}
and the action variables $w_k$
\begin{gather}\label{w-st}
\omega_i=\dfrac{1}{n}\sum_{j=1}^n\int^{q_j} \dfrac{S_{ij}(\lambda)}{\sqrt{\sum\limits_{k=1}^n H_k S_{kj}(\lambda)-V_j(\lambda)}}\,\mathrm d\lambda,
\end{gather}
 so that
\begin{gather*}
\{H_j,H_k\}=\{\omega_i,\omega_k\}=0,\qquad \{H_j,\omega_k\}=\delta_{ij}.
\end{gather*}

In generic case the action variables (\ref{w-st}) are sums of the multi-valued  Abelian integrals. However, if we are able to apply the known addition theorems for these Abelian integrals then we can get additional integrals of motion \cite{ts08a,ts09e, ts08m}.

Let us discuss addition theorems for the logarithmic (exponential) and elliptic functions  \cite{bak97,eul68}. In the f\/irst case polynomials
\[
P_j=\sum_{k=1}^n H_k S_{kj}(\lambda)-V_j(\lambda)
\]
are the second-order polynomials and
\[
\omega=\dfrac{1}{n}\sum_{j=1}^m\int^{q_j} \dfrac{d \lambda}{\sqrt{k_j^2\lambda^2+b_j\lambda+c_j}}=
\sum_{j=1}^m
\ln\left(p_j+\dfrac{k_j^2q_j+b_j}{k_j}\right)^{\frac{1}{nk_j}} .
\]
So, we can easily get polynomial or rational function
\begin{gather}\label{add-th1}
Z=e^{z\omega}=\prod_{j=1}^m\left(p_j+\dfrac{k_j^2q_j+b_j}{k_j}\right)^{\frac{z}{nk_j}}
\end{gather}
at the special choice of $k_j$ and $z$. In the second case  $m-1$ angle variables $\omega_k,\ldots,\omega_{k+m-1}$ with $1<m\leq n$ have to  satisfy to the Abel equations
\begin{gather*}
 \dfrac{d\mathrm x_1}{\sqrt{f(\mathrm x_1)}}+\dfrac{d\mathrm x_2}{\sqrt{f(\mathrm x_2)}}+\cdots+\dfrac{d\mathrm x_m}{\sqrt{f(\mathrm x_m)}}=d\omega_k, \\
 \dfrac{\mathrm x_1d\mathrm x_1}{\sqrt{f(\mathrm x_1)}}+\dfrac{\mathrm x_2d\mathrm x_2}{\sqrt{f(\mathrm x_2)}}+\cdots+\dfrac{\mathrm x_md\mathrm x_m}{\sqrt{f(\mathrm x_m)}}=d\omega_{k+1}, \\
 \cdots\cdots\cdots\cdots\cdots\cdots\cdots\cdots\cdots\cdots\cdots\cdots\cdots\cdots\cdots \\ 
 \dfrac{\mathrm x_1^{m-2}d\mathrm x_1}{\sqrt{f(\mathrm x_1)}}+\dfrac{\mathrm x_2^{m-2}d\mathrm x_2}{\sqrt{f(\mathrm x_2)}}+\cdots+\dfrac{\mathrm x_m^{m-2}d\mathrm x_m}{\sqrt{f(\mathrm x_m)}}=d\omega_{k+m-1} ,
\end{gather*}
with a common polynomial of f\/ixed degree $2m$
\begin{gather}\label{h-curve}
f(\mathrm x)\equiv A_{2m}\mathrm x^{2m}+A_{2m-1}\mathrm x^{2m-1}+\cdots+A_1 \mathrm x+A_0.
\end{gather}
If this case there are some additional Richelot integrals of motion \cite{rich42}
\begin{gather*}
C_k=\dfrac{\left[\dfrac{\sqrt{f(\mathrm x_1)}}{F'(\mathrm x_1)}\cdot\dfrac{1}{a_k-\mathrm x_1}+\cdots+\dfrac{\sqrt{f(\mathrm x_m)}}{F'(\mathrm x_m)}\cdot\dfrac{1}{a_k-\mathrm x_m}\right]^2}
{\left[\dfrac{\sqrt{f(\mathrm x_1)}}{F'(\mathrm x_1)}+\cdots+\dfrac{\sqrt{f(\mathrm x_m)}}{F'(\mathrm x_m)}\right]^2-A_{2m}} F(a_k) .
\end{gather*}
Here $a_k$ are  values of $\mathrm x$ at the branch points of the corresponding hyperelliptic curve  and $F(\mathrm x)=(\mathrm x-\mathrm x_1)(\mathrm x-\mathrm x_2)\cdots(\mathrm x-\mathrm x_m)$ \cite{rich42}.

At $m=2$ we have famous Euler algebraic integral  \cite{eul68}. If  $A_{2m}=0$ and $A_{2m-1}\neq 0$ in~(\ref{h-curve}) there is another additional Richelot integrals of motion \cite{rich42}.   The Weierstrass generating function of such integrals for any values of the coef\/f\/icients  $A_k$ and other constructions of  additional integrals of the Abel equations are discussed in~\cite{bak97}. Some example of the Euler and Richelot superintegrable systems  may be found in \cite{ts09a,ts08a,ts10r}.

Below we show how these addition theorems  could help us to classify  superintegrable systems.

\section{Elliptic coordinate system}

Let us consider elliptic coordinates on the plane $q_{1,2}$ def\/ined by
\[
1-\dfrac{x^2}{\lambda-\kappa}-\dfrac{y^2}{\lambda+\kappa}=\dfrac{(\lambda-q_1)(\lambda-q_2)}{\lambda^2-\kappa^2} ,\qquad \kappa\in \mathbb R .
\]
The corresponding momenta reads as
\[
p_1 = \dfrac{2p_xx}{q_1-\kappa}+\dfrac{2p_yy}{q_1+\kappa} ,\qquad
p_2 =\dfrac{2p_xx}{q_2-\kappa}+\dfrac{2p_yy}{q_2+\kappa} .
\]
The St\"{a}ckel matrix and the separated relations
\begin{gather*}
S=
   \begin{pmatrix}
    \dfrac{q_1}{q_1^2-\kappa^2} & \dfrac{q_2}{q_2^2-\kappa^2} \vspace{2mm}\\
          \dfrac{1}{q_1^2-\kappa^2} & \dfrac{1}{q_2^2-\kappa^2} \\
   \end{pmatrix},\qquad \begin{array}{@{}l}
                       p_1^2+V_1-\dfrac{q_1H_1}{q_1^2-\kappa^2}-\dfrac{H_2}{q_1^2-\kappa^2}=0 ,
                    \vspace{2mm}\\
                    p_2^2+V_2-\dfrac{q_2H_1}{q_2^2-\kappa^2}-\dfrac{H_2}{q_2^2-\kappa^2}=0 ,
                    \end{array}
\end{gather*}
give rise to the following  Hamiltonians in the involution
\begin{gather*}
H_1 = \dfrac{(q_1^2-\kappa^2)(p_1^2+V_1)}{q_2-q_1}+\dfrac{(q_2^2-\kappa^2)(p_2^2+V_2)}{q_2-q_1} , \\
H_2 = \dfrac{q_2(q_1^2-\kappa^2)(p_1^2+V_1)}{q_1-q_2}-\dfrac{q_1(q_2^2-\kappa^2)(p_2^2+V_2)}{q_1-q_2} .
\end{gather*}
The Hamiltonian $H_1$ commutes with the second angle variable $w_2$, which is equal to
\begin{gather}
w_2 = \dfrac{1}{2}\int^{q_1} \dfrac{d\lambda}{\sqrt{(\lambda^2-\kappa^2)(\lambda H_1+H_2-V_1\lambda^2+V_1\kappa^2)}} \nonumber\\
\phantom{w_2 =}{}  + \dfrac{1}{2}\int^{q_2} \dfrac{d\lambda}{\sqrt{(\lambda^2-\kappa^2)(\lambda H_1+H_2-V_2\lambda^2+V_2\kappa^2)}} . \label{ell-w}
\end{gather}
Polynomials
\begin{gather}\label{quart-ell}
P_{1,2}=\big(\lambda^2-\kappa^2\big)\big(\lambda H_1+H_2-V_{1,2}\lambda^2+V_{1,2}\kappa^2\big)
\end{gather}
standing under square root in these integrals are at least third-order polynomials on $\lambda$. So,  in this case we can not apply addition theorem for the logarithms.

It is easy to see that we can apply addition theorem  for the elliptic functions at
\[
V_1=V_1=\alpha .
\]
Namely, if we put  $\lambda=\mathrm x$ and $\lambda=\mathrm y$ in the f\/irst and second integrals (\ref{ell-w}),  we could apply  the Euler addition theorem
\begin{gather}\label{ell-add}
\dfrac{\mathrm{dx}}{\sqrt{\mathrm X}}+\dfrac{\mathrm{dy}}{\sqrt{\mathrm Y}}=\dfrac{\mathrm{ds}}{\sqrt{\mathrm S}}
\end{gather}
to  angle variable $\omega_2$. Here $\mathrm X$ is an arbitrary quartic
\begin{gather}\label{X-quart}
\mathrm X = a\mathrm x^4 + 4b\mathrm x^3 + 6c\mathrm x^2 +4d\mathrm x + e
\end{gather}
and  $\mathrm Y$, $\mathrm S$ are the same functions  of another variables $\mathrm y$ and $\mathrm s$.
In this case, symmetrical  biquadratic form of $\mathrm x$ and $\mathrm y$
\begin{gather*}
F(\mathrm x,\mathrm y)=a \mathrm x^2\mathrm y^2+2b\mathrm x \mathrm y(\mathrm x+\mathrm y)+c\big(\mathrm x^2+ 4\mathrm x\mathrm y +\mathrm y^2\big)+2d(\mathrm x+\mathrm y)+e=0
\end{gather*}
def\/ines the conic section on the plane $(\mathrm x,\mathrm y)$. According to \cite{bak97, eul68,rich42},  there is a famous Euler integral
\begin{gather}\label{s-small}
C=\dfrac{F(\mathrm x,\mathrm y)-\sqrt{\mathrm X}\sqrt{\mathrm Y}}{2(\mathrm x-\mathrm y)^2}=
\dfrac{1}{4}\left(\dfrac{\sqrt{\mathrm X}-\sqrt{\mathrm Y}}{\mathrm x-\mathrm y}\right)^2- \dfrac{a(\mathrm x+\mathrm y)^2}4-b(\mathrm x+\mathrm y)-c.
\end{gather}
For the quartic (\ref{quart-ell}) associated with the angle variable (\ref{ell-w})  this Euler integral looks like
\[
H_3=\dfrac{(p_1-p_2)(q_1^2-\kappa^2)(q_2^2-\kappa^2)}{(q_1-q_2)^3}
\Bigl(\alpha (q_1-q_2)^2-(p_1-p_2)^2\kappa^2+(p_1q_1-p_2q_2)^2\Bigr) .
\]
Here $q_{1,2}$ and $p_{1,2}$ are  elliptic  coordinates and momenta.

It is a third-order polynomial in momenta which commutes with the Hamiltonian
\[
\{H_1,H_2\}=0 ,\qquad \{H_2,H_3\}=H_4\neq 0 .
\]
The algebra of the polynomial integrals of motion $H_1$, $H_2$, $H_3$ can be closed only after some other polynomial generators are added.

Thus,  we easily  f\/ind  the additional integrals of motion for the Hamilton function of the oscillator
\[
4H_1=p_x^2+p_y^2+a\big(x^2+y^2\big)
\]
using the separation of variables in elliptic coordinate system and the corresponding angle va\-riab\-les.
Another result is that there is only one superintegrable system separable in elliptic coordinates and associated with the known addition theorems for Abelian integrals.

\section{Parabolic coordinate system}

Let us consider parabolic coordinates on the plane $q_{1,2}$ def\/ined by
\[
x = q_1q_2,\qquad y = \dfrac{q_1^2-q_2^2}{2}
\]
and the corresponding momenta
\[
p_x = \dfrac{p_1q_2+p_2q_1}{q_1^2+q_2^2} ,\qquad
p_y = \dfrac{q_1p_1-q_2p_2}{q_1^2+q_2^2} ,
\]
The St\"{a}ckel matrix and the separated relations
\begin{gather}
S=
   \begin{pmatrix}
     q_1^2 & q_2^2 \\
     1 & -1 \\
   \end{pmatrix}, \qquad \begin{array}{@{}l}
p_1^2+V_1(q_1)-q_1^2H_1-H_2=0 ,\vspace{1mm}\\
                                     p_2^2+V_2(q_2)-q_2^2H_1+H_2=0,
\end{array}\label{par-srel}
\end{gather}
give rise to the Hamiltonians
\[
H_1 =\dfrac{p_1^2+p_2^+V_1(q_1)+V_2(q_2)}{q_1^2+q_2^2} ,\qquad
H_2 =\dfrac{ p_1^2q_2^2-p_2^2q_1^2+q_2^2V_1(q_1)-q_1^2V_2(q_2)}{q_1^2+q_2^2} .
\]
The Hamiltonian $H_1$ commutes with the second angle variable $w_2$, which is equal to
\[
w_2=\dfrac{1}{2}\int^{q_1} \dfrac{d\lambda}{\sqrt{\lambda^2 H_1+H_2-V_1(\lambda)}} +\dfrac{1}{2}\int^{q_2} \dfrac{d\lambda}{\sqrt{\lambda^2 H_1-H_2-V_2(\lambda)}} .
\]
In contrast with the elliptic coordinates, polynomials
\[
P_{1,2}=\lambda^2 H_1+H_2-V_{1,2}(\lambda)
\]
standing  under square root in these integrals are at least second-order polynomials on~$\lambda$. So, we can apply both known addition theorems to these Abelian integrals.

In fact, these integrals are expressed via logarithmic functions if\/f:
\begin{gather*}
   \mbox{Case 1}:\quad  V_1=b_1q_1+c_1 ,\qquad V_2=b_2q_2+c_2 ,\\
 \mbox{Case 2}:\quad  V_1=a_1q_1^{-2}+b_1 ,\qquad V_2=a_2q_2^{-2}+b_1 .
\end{gather*}
The addition theorem for the elliptic function is applicable if\/f:
\begin{gather*}
\mbox{Case 3}:\quad V_1=a_1q_1^6+b_1q_1^4+c_1q_1^{-2} ,\qquad
V_2=a_1q_2^6-b_1q_2^4+c_1q_2^{-2} .
\end{gather*}
The corresponding Hamilton functions  are deformations of the Kepler--Coulomb  and oscillator Hamiltonians:
\begin{gather*}
  \mbox{Case 1}:\quad
H_1 = p_x^2+p_y^2+\frac{1}{2\sqrt{x^2+y^2}}\left(b_1 \sqrt{x+\!\sqrt{x^2+y^2}}+b_2 \sqrt{\!\sqrt{x^2+y^2}-x}+c_1+c_2\right)
\\
\hphantom{\mbox{Case 1}:\quad
H_1 =}{} = p_x^2+p_y^2+ \frac{1}{2r}\left(b_1 \sqrt{2} \cos \frac \varphi 2+b_2 \sqrt{2} \sin \frac \varphi 2+c_1+c_2\right);
\\
\mbox{Case 2}:\quad
H_1 = p_x^2+p_y^2+
\frac1{2\sqrt{x^2+y^2}}\left(\frac{a_1}{{x+\sqrt{x^2+y^2}}}+\frac{a_2}{{x-\sqrt{x^2+y^2}}}+b_1+b_2\right) \\
\hphantom{\mbox{Case 2}:\quad
H_1 =}{}
= p_x^2+p_y^2+ {\frac {a_1}{2{r}^{2} \left( \cos  \varphi  +1
 \right) }}-{\frac {a_2}{2{r}^{2} \left( \cos  \varphi
  -1 \right) }}+{\frac {b_1+b_2}{2r}};
\\
\mbox{Case 3}:\quad
H_1 = p_x^2+p_y^2+{\alpha}\left( 4 {x}^{2}+{y}^{2} \right) +2 \beta x+{\frac {
\gamma}{{y}^{2}}} .
\end{gather*}
Here $r=\sqrt{x^2+y^2}$ and $\varphi=\arctan x/y$ are polar coordinates on the plane. According to \cite{w12} these systems remain superintegrable in the quantum case.

\subsection{Case 1}
In the f\/irst case the second angle variable equals
\begin{gather*}
\omega_2 = \dfrac{1}{2}\int^{q_1}\dfrac{d\lambda}{\sqrt{\lambda^2H_1-b_1\lambda+H_2-c_1}}-
\dfrac{1}{2}\int^{q_2}\dfrac{d\lambda}{\sqrt{\lambda^2H_1-b_2\lambda-H_2-c_2}}
 \\
\phantom{\omega_2}{}  =
\dfrac{\ln\left(p_1-\frac{q_1H_1-b_1/2}{\sqrt{H_1}}\right)}{2\sqrt{H_1}}
-\dfrac{\ln\left(p_2-\frac{q_2H_1-b_2/2}{\sqrt{H_1}}\right)}{2\sqrt{H_1}} .
\end{gather*}
 The application of the addition theorem (\ref{add-th1}) to  $\omega_2$ gives rise to the following  rational integral of motion
\[
Z=e^{2\sqrt{H_1}\omega_2}=\dfrac{2q_1H_1+2p_1\sqrt{H_1}-b_1}{2q_2H_1+2p_2\sqrt{H_1}-b_2} .
\]
In order to calculate polynomial integral of motion let us consider a series expansion of the function
\[
f=\dfrac{1}{\sqrt{H_1}} \left(\alpha Z+\beta Z^{-1}\right)=\dfrac{\alpha(4H_1H_2-4H_1c_1-b_1^2)-\beta(4H_1H_2+4H_1c_2+b_2^2)}
{\sqrt{H_1(4H_1c_2+4H_1H_2+b_2^2)(b_1^2+4H_1c_1-4H_1H_2)}}+O(p_{1,2}) ,
\]
 by momenta $p_{1,2}$. Here we substitute the variables  $q_{1,2}$ from the separated relations (\ref{par-srel}) into the rational integral $Z$  and  $\alpha$, $\beta$ are undef\/ined polynomials in~$H_{1,2}$.

 Equating f\/irst coef\/f\/icient of this expansion to zero one gets the following expressions for these polynomials
 \[
\alpha= 4H_1H_2+4H_1c_2+b_2^2 ,\qquad \beta=4H_1H_2-4H_1c_1-b_1^2 .
 \]
 At this values of $\alpha$ and $\beta$ the function $f$ becomes  a third-order polynomial in momenta
 \begin{gather*}
 H_3 = \dfrac{1}{\sqrt{H_1}} \left(\alpha Z+\beta Z^{-1}\right)=\dfrac{1}{\sqrt{H_1}} \left(\alpha e^{2\sqrt{H_1}\omega_2}+\beta e^{-2\sqrt{H_1}\omega_2}\right) \\
\hphantom{H_3}{}
=\frac{8(p_1q_2-p_2q_1)(p_1^2+p_2^2+c_1+c_2)}{q_1^2+q_2^2}
+\frac{4\bigl(2p_1q_1q_2-p_2(q_1^2-q_2^2)\bigr)b_1}{q_1^2+q_2^2}\\
\hphantom{H_3=}{}
-\frac{4\bigl(2p_2q_1q_2+p_1(q_1^2-q_2^2)\bigr)b_2}{q_1^2+q_2^2} ,
  \end{gather*}
  such that
  \[
  \{H_1,H_3\}=0 .
  \]
In order to close the algebra of the polynomial integrals of motion $H_1$, $H_2$, $H_3$ we have to add one more polynomial generator
\[
H_4=\{H_2,H_3\}=2\alpha e^{2\sqrt{H_1}\omega_2}-2\beta e^{-2\sqrt{H_1}\omega_2} .
\]
by analogy with $\exp(\omega)$, $\sin(\omega)$ and $\cos(\omega)$ functions.
\begin{remark}
One of the referees proposed another construction of the polynomial integrals of motion from the angle variable $\omega_2$. Namely,  from the separation equations we can deduce that
\begin{gather*}
\Delta_1=(2q_1H_1+2p_1\sqrt{H_1}-b_1)(2q_1H_1-2p_1\sqrt{H_1}-b_1)=-4H_1H_2+4H_1c_1+b_1^2 ,\\
\Delta_2=(2q_2H_1+2p_2\sqrt{H_1}-b_2)(2q_2H_1-2p_2\sqrt{H_1}-b_2)=4H_1H_2+4H_1c_2+b_2^2 .
\end{gather*}
We can therefore write
\begin{gather*}
 \Psi_1=\Delta_1Z^{-1} = \bigl( (2q_1H_1-b_1)(2q_2H_2-b_2)-4H_1p_1p_2\bigr) \\
\phantom{\Psi_1=\Delta_1Z^{-1} =}{} - 2\sqrt{H_1}\bigl((2q_2H_1-b_2)p_1-(2q_1H_1-b_1)p_2 \bigr)
\end{gather*}
and
\begin{gather*}
 \Psi_2=\Delta_2Z = \bigl( (2q_1H_1-b_1)(2q_2H_2-b_2)-4H_1p_1p_2\bigr) \\
\phantom{\Psi_2=\Delta_2Z =}{} + 2\sqrt{H_1}\bigl((2q_2H_1-b_2)p_1-(2q_1H_1-b_1)p_2 \bigr) .
\end{gather*}
Consequently
\[
H_3=\dfrac{1}{\sqrt{H_1}}(\Psi_1-\Psi_2)=-4\bigl((2q_2H_1-b_2)p_1-(2q_1H_1-b_1)p_2 \bigr)
\]
is a third-order constant of motion. This method of explanation may be  clearer than the method of expansion of the rational in momenta function $f=\alpha Z+\beta Z^{-1}$ with  indef\/inite coef\/f\/icients $\alpha$ and $\beta$.
\end{remark}

\begin{remark}
Let us remind, that two-dimensional open  Toda lattice def\/ined by the following polynomial integrals of motion
\[
H_1=p_1^2+p_2^2+\e^{q_1-q_2} ,\qquad H_2=p_1+p_2 ,
\]
has the non-rational in momenta additional integral of motion
\[
Z=\dfrac{p_1-p_2+\sqrt{J}}{p_1-p_2-\sqrt{J}}
\exp\left(\sqrt{J} \dfrac{q_1+q_2}{p_1+p_2}\right) ,\qquad J=2H_2-H_1^2 .
\]
which can be also obtained from the second angle variable~$\omega_2$~\cite{ts08m}.  However, it is easy to prove,  that we can not apply  working above  constructions of polynomial integral of motion $H_3$ in this case.
\end{remark}

\subsection{Case 2}

In the second case the angle variable is equal to
\[
\omega_2=\dfrac{1}{2}\int^{q_1}\dfrac{\lambda d\lambda}{\sqrt{\lambda^4 H_1+(H_2-b_1)\lambda^2-a_1}} -\dfrac{1}{2}\int^{q_2}\dfrac{\lambda d\lambda}{\sqrt{\lambda^4 H_1-(H_2+b_2)\lambda^2-a_2}} .
\]
Changing variables $\mu=\lambda^2$ one gets second-order polynomials
 \[
 P_j=\mu^2 H_1\pm(H_2\mp b_j) \mu-a_j
 \]
 under the square root and desired sum of the logarithms
\[
\omega_2=\dfrac{\ln\left(2q_1^2H_1+2q_1p_1\sqrt{H_1}+H_2-b_1)\right)}{4\sqrt{H_1}}
-\dfrac{\ln\left(2q_2^2H_1+2q_2p_2\sqrt{H_1}-H_2-b_2)\right)}{4\sqrt{H_1}} .
\]
The rational integral of motion (\ref{add-th1}) is equal to
\[
Z=e^{4\sqrt{H_1}\omega_2}=\dfrac{2q_1^2H_1+2q_1p_1\sqrt{H_1}+H_2-b_1}{2q_2^2H_1+2q_2p_2\sqrt{H_1}-H_2-b_2} .
\]
As above we consider the expansion  of the function
\[
f=\dfrac{1}{\sqrt{H_1}} \left(\alpha Z+\beta Z^{-1}\right)
\]
by momenta $p_{1,2}$.  Equating f\/irst coef\/f\/icient of this expansion to zero one gets polynomials~$\alpha$,~$\beta$
\[
\alpha=4H_1a_2+b_2^2+2b_2H_2+H_2^2 ,\qquad
\beta=-4H_1a_1-b_1^2+2b_1H_2-H_2^2 .
\]
 At this values of $\alpha$ and $\beta$  the function $f$ becomes  a third-order polynomial in momenta
 \begin{gather*}
 H_3 =\dfrac{1}{\sqrt{H_1}} \left(\alpha Z+\beta Z^{-1}\right)=\dfrac{1}{\sqrt{H_1}}
 \left(\alpha e^{4\sqrt{H_1}\omega_2}+\beta e^{-4\sqrt{H_1}\omega_2}\right)\\
\hphantom{H_3}{} =\frac{4(q_1p_1-q_2p_2)(q_2p_1-q_1p_2)^2}{q_1^2+q_2^2}
 + \frac{4q_1q_2(q_2p_1-q_1p_2)(b_1+b_2)}{q_1^2+q_2^2}\\
\hphantom{H_3=}{}
+\frac{4a_1q_2\bigl(q_2q_1p_1-(2q_1^2+q_2^2)p_2\bigr)}{q_1^2(q_1^2+q_2^2)}
-\frac{4a_2q_1\bigl(q_2q_1p_2-(2q_2^2+q_1^2)p_1\bigr)}{q_2^2(q_1^2+q_2^2)} .
\end{gather*}
 such that
  \[
  \{H_1,H_3\}=0 .
  \]
In order to close the algebra of the polynomial integrals of motion $H_1$, $H_2$, $H_3$ we have to add one more polynomial generator
\[
H_4=\{H_2,H_3\}=4\alpha e^{4\sqrt{H_1}\omega_2}-4\beta e^{-4\sqrt{H_1}\omega_2} .
\]

\subsection{Case 3}
In the third case the  angle variable is equal to
\begin{gather*}
\omega_2=
\int^{q_1}\!\!\dfrac{\lambda d\lambda}{\sqrt{-a_1\lambda^8-b_1\lambda^6+H_1\lambda^4+H_2\lambda^2-c_1}}
-\int^{q_2}\!\!\dfrac{\lambda d\lambda}{\sqrt{-a_1\lambda^8+b_1\lambda^6+H_1\lambda^4-H_2\lambda^2-c_1}} .\!
\end{gather*}
Changing variables $\lambda=\sqrt{\mathrm x}$ and $\lambda=i\sqrt{\mathrm y}$ at the f\/irst and second integral one gets the Euler addition theorem~(\ref{ell-add}).  In fact, this example has been considered in Euler's book~\cite{eul68} too.

Identifying quartic
\[
P=-a_1\mu^4-b_1\mu^3+H_1\mu^2+H_2\mu-c_1
\]
with $\mathrm X$ (\ref{X-quart}) we can easily calculate the Euler  integral of motion (\ref{s-small}) in parabolic coordinates
 \begin{gather*}
H_3=s =\frac{(q_1p_1-q_2p_2)(q_1p_2+q_2p_1)^2}{(q_1^2+q_2^2)^3 }
+\frac{a_1q_1q_2(2q_1^3p_2+q_2q_1^2p_1-q_1q_2^2p_2-2q_2^3p_1)}{q_1^2+q_2^2}\\
\hphantom{H_3=s =}{}
+\frac{b_1q_1q_2(q_1p_2+q_2p_1)}{q_1^2+q_2^2}
 +\frac{c_1(q_1p_1-q_2p_2)}{q_1^2q_2^2(q_1^2+q_2^2)} .
\end{gather*}
The algebra of the integrals of motion $H_1$, $H_2$, $H_3$ is more complicated then the algebra associated with the addition theorem for logarithms. In fact, in order to close this algebra we have to introduce the counterparts of the Jacobi elliptic functions sn($\omega$), cn($\omega$) and dn($\omega$) instead of the  trigonometric functions $\sin(\omega)$ and $\cos(\omega)$, which we used for the superintegrable systems associated with the addition theorem for logarithms.

\section{Conclusion}
It is known that orthogonal coordinate systems on Riemaniann manifolds can be viewed as an orthogonal
sum of certain basic coordinate systems  and  these basic  systems  can be obtained from the elliptic coordinate system \cite{kal} using a degeneration procedure. This degeneration decreases the degree of
  polynomials standing  under square roots into the angle variables~(\ref{w-st}).  Thus,   we have only one superintegrable systems separable in elliptic coordinates, whereas for degenerations we have a lot of dif\/ferent   superintegrable systems.  As usual, the addition theorem for logarithms allows us to get additional integrals of higher order in momenta~\cite{ts08a,ts09e}.

\subsection*{Acknowledgments}

We are greatly indebted referees for several
improvements and corrections induced by their comments.

\pdfbookmark[1]{References}{ref}
\LastPageEnding

\end{document}